\documentclass[usenatbib,useAMS]{mn2e}
\usepackage{graphicx}

\usepackage{aas_macros}
\title[]{Rotation of young stars in Cepheus OB3b}

\author[]{S.\,P.\ Littlefair$^{1}$, Tim Naylor$^{2}$, N.\,J.\ Mayne$^{2}$, Eric S.\,Saunders$^{3}$, 
R.\,D\ Jeffries$^{4}$ \\
$^1$Dept of Physics and Astronomy, University of Sheffield, S3 7RH, UK \\
$^2$School of Physics, University of Exeter, Exeter, EX4 4QL, UK\\
$^3$Las Cumbres Observatory, 6740 Cortona Dr., Suite 102, Santa Barbara, CA 93117\\
$^4$School of Chemistry and Physics, Keele University, Keele, Staffordshire, ST5 5BG, UK\\}

\date{\center{\Large Submitted for publication in the Monthly Notices of the
Royal Astronomical Society \\ 
\vspace{.5cm} \today}} 

\begin{document}
\maketitle

\begin{abstract} 
We present a photometric study of $I$-band variability in the young association Cepheus OB3b. The study is sensitive 
to periodic variability on timescales of less than a day, to more than 20 days. After rejection of contaminating objects 
using $V$, $I$, $R$ and narrowband H$\alpha$ photometry, we find 475 objects with measured rotation periods, 
which are very likely pre-main-sequence members of the Cep OB3b star forming region. 

We revise the distance and age to Cep OB3b, putting it on the self-consistent age and distance ladder of 
\cite{2008MNRAS.386..261M}. This yields a distance modulus of 8.8$\pm$0.2 mags, corresponding to a distance of 
580$\pm$60 pc, and an age of 4-5Myrs. 

The rotation period distribution confirms the general picture of rotational evolution in young stars, exhibiting both the 
correlation between accretion (determined in this case through narrowband H$\alpha$ photometry) and rotation 
expected from disc locking, and the dependence of rotation upon mass that is seen in other star forming regions. 
However, this mass dependence is much weaker in our data than found in other studies. Comparison to the similarly 
aged NGC 2362 shows that the low-mass stars in Cep OB3b are rotating much more slowly. This points to a possible 
link between star forming environment and rotation properties. Such a link would call into question models of stellar 
angular momentum evolution, which assume that the rotational period distributions of young clusters and 
associations can be assembled into an evolutionary sequence, thus ignoring environmental effects.
\end{abstract} 


\section{Introduction}
\label{sec:introduction}

There are sound theoretical reasons to expect that accretion processes
help determine the angular momentum of T-Tauri
stars. Historically, the slow rotation rates of T Tauri stars
(relative to their break-up velocity) has been explained by the
disc-locking theory \citep{konigl91,shu94}. In this theory, magnetic
field lines connect the star to the disc, enforcing synchronous
rotation between the star and the material in the disc at some radius,
near where the magnetic field disrupts the disc. The simplistic theory has
been expanded in a variety of models where angular momentum is removed
from the star by a combination of the disc and an accretion-driven 
wind \citep[e.g.][]{fendt07,romanova07,matt08}. Whether the star is spun up or down
by the star-disc interaction depends upon the balance between accretion spinning
up the star and magnetic (or wind) torques slowing it down. Theoretically, this issue is
unresolved; some studies find the star spins up \citep{bessolaz08}, some find it spins
down \citep[e.g.][]{long05}.

Observationally, the evidence for the influence of accretion disks on rotation is much stronger than
it was a few years ago \citep{herbst07}. Previously, conflicting results had arisen \citep[e.g.][]
{herbst02,stassun99,littlefair05}. These were most likely due to a combination of small sample sizes, and ambiguous 
diagnostics of the presence of accretion disks. \cite{rebull06} resolved these issues with a large sample of rotation 
periods in the ONC, and accretion disk status defined from Spitzer IRAC data.
\cite{rebull06} found a clear correlation between mid-IR excess and rotation, in the sense that stars 
with mid-IR excess were much more likely to be slow-rotators. An analysis of the slightly older NGC 
2264 found the same result \citep{cieza07}, and confirmed the \cite{rebull06} result, via a refined 
analysis of the same data. There is thus strong observational evidence that the star-disc interaction is  
responsible for extracting angular momentum from young stars.
Interestingly, a small population of rapidly rotating stars with mid-IR excesses exists; it is possible that 
these stars are being spun-up by accretion, hinting that the braking process of young stars is 
intermittent. Also, in both the ONC and NGC 2264, there exists a significant population of slow 
rotators with {\em no} mid-IR excess. These are often interpreted as being recently released from 
disc-locking, but there are problems with this interpretation \citep[see][for example]{bouvier07}.

The firm link established between accretion and rotation represents significant progress, but there are 
still open questions regarding the rotation of young stars. For example, the low-mass stars are 
rotating more rapidly than the high-mass stars, and appear to spin-up more rapidly as they contract 
towards the main sequence \citep{herbst02,irwin07a}. The reason for this is still not known. Also, the 
angular momentum evolution of young stars is determined by assembling different clusters into an 
evolutionary sequence, and assuming the period distribution of the older clusters can be modelled 
using the young clusters as a starting point. In doing so, the possibility of an environmental effect on 
rotation is ignored. Such an environmental effect may be indicated by the data; the young stars in 
IC348 rotate much slower than those in the similarly aged NGC 2264 \citep{littlefair05}.

Here we present a photometric study of the young association Cepheus OB3b (hereafter Cep OB3b). Cep OB3 is a 
young association covering a region of the sky from approximately 22$^h$46$^m$ to 23$^h$10$^m$ in right 
ascension and +61$^{\circ}$ to +64$^{\circ}$ in declination. The subgroup Cep OB3b lies closest to the molecular 
cloud, and has a rich pre-main-sequence (PMS) population, confirmed by both spectroscopy \citep{pozzo03} and X-
ray data \citep{getman06}. In section~\ref{sec:obs} we present the observations and data reduction techniques 
applied. 
Section~\ref{sec:detect} describes the techniques used to identify periodic variables. Section~\ref{sec:distance} 
provides a revision of the age of, and distance to, Cep OB3b. In Sections~\ref{sec:results} and \ref{sec:discussion} we 
present our results and analysis of the data, whilst in section~\ref{sec:concl} we draw our conclusions.

\section{Observations and Data Reduction}
\label{sec:obs}

\subsection{Observations}
\label{subsec:obs}
RGO $I$-band CCD images were taken with the Wide Field Camera on the
Isaac Newton Telescope (INT) on La Palma, equipped with 4 thinned EEV
2kx4k CCDs. The total sky coverage per exposure is 0.29 sq deg. A
single field, with the centre of CCD \#4 pointing at $\alpha = 22^h
55^m 43.3^s$, $\delta = +62^d 40^m 13^s$ J2000, was observed, with
data being taken on every night between 21st September 2004 and
6th October 2004.  This dataset will be referred to as the "short baseline" (SB) dataset.
Because the SB dataset is not sensitive to periods longer than 7--10 days, we augmented it with two 
additional datasets, both of which have longer baselines, but less dense sampling.
Additional data was taken with the wide field camera on 29 nights between 23rd August 2005 and 1st 
November 2005. This dataset is referred to as the "long baseline 0" (LB0) dataset.

For the SB dataset the seeing varied between 0.8 and 5\arcsec, with a median and standard 
deviation of 1.1\arcsec and 0.9\arcsec respectively.  Most nights were affected by thin to heavy cirrus 
cloud, although the nights between on 26th September 2004 and 29th September 2004, and the 
night of the 5th October 2004 were photometric.  
The LB0 dataset was taken in seeing conditions between 0.8 and 2.6\arcsec, with a median and 
standard deviation of 0.9\arcsec and 0.8\arcsec respectively. Most nights were affected by light cirrus 
only, with less than half a magnitude effect on transparency. To increase the dynamic range of the SB 
and LB0 datasets, we used exposure times of 5, 30 and 300 seconds. For the SB dataset these were 
repeated many times throughout a night; for the LB0 dataset one or two exposures were taken each night.  
In total the SB dataset contains 477 useable short  exposures, 617 useable medium exposures and 614 useable long exposures.  The LB0 dataset contains 41 useable short  exposures, 41 useable medium exposures and 36 useable long exposures. 

The second long-baseline dataset consists of SDSS $i'$-band CCD images obtained with DillCam on 
the 2-m Faulkes Telescope N (FTN). DillCam has a single CCD with a pixel scale of 0.278 arcsec per 
pixel in the default 2$\times$2 binning mode, giving a 4.7$\times$4.7 arcmin field of view. Four fields 
were observed, chosen to cover the positions of highest stellar density. Data were taken on 15 nights between 13th October 2005 and 17th December 2005, with between 2--4 separate observations per night. Exposure times of 30 and 300 seconds were used. The number of exposures varies from field to field, and is different for the short and long exposure times, but ranges from 43 to 60 exposures in total. This dataset is referred to as the "long baseline 1" (LB1) dataset.  The LB1 dataset was taken in seeing conditions between 1.0 and 3.1\arcsec, with a median and standard 
deviation of 1.3\arcsec and 0.6\arcsec respectively. The data are only marginally affected by cloud, 
with transparency variations always less than 0.3 mags.

\begin{figure}
\includegraphics[scale=0.37,angle=0,trim=0 0 0 0]{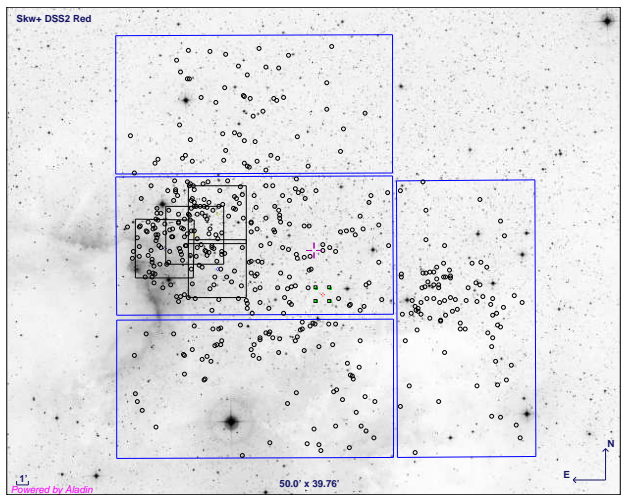}
\caption{INT WFC field-of-view overlayed on DSS2 red image of Cep OB3. Also shown are the four 
Faulkes telescope pointings obtained. Circles denote the locations of periodic association members 
(see section~\ref{sec:detect} for details).}
\label{fig:pointing}
\end{figure}

A plot of the Cep OB3 region and the fields observed is shown in figure~\ref{fig:pointing}.

Over the nights of September 12th, 13th and 14th 2003, we used the INT to obtain three sets of images of the same field as is covered by our $I$-band data, each set consisting of a 300s and a 30s exposure in H$\alpha$
and a 30s and a 3s exposure in $R$.  We debiased and flatfielded the images in a similar manner to the $I$-band 
data, but then instead of searching for stars in a combined image, we carried out optimal photometry at the positions 
of all the objects in the catalogue of \cite{mayne07}. See Section~\ref{subsec:phot} for a complete description of how 
the photometry was carried out. We then profile corrected and combined the magnitudes from each image,  as 
described in \cite{mayne07} , to obtain $R$ and $R-H\alpha$ magnitudes, with an arbitrary zero point.  We then 
combined this with the V and I photometry of \cite{mayne07} to create Table 1 (the full version of which is available electronically).

\begin{table*}
\begin{center}
\caption[]{Photometry for the INT WFC field observed is available in this format at http://www.astro.ex.ac.uk/people/timn/Catalogues/tables.html. This table is a sample only, to indicate the format of the online table. The flag column indicates data quality, and is explained in full at http://www.astro.ex.ac.uk/people/timn/Catalogues/format.html}
\begin{tabular}{@{\extracolsep{-1.25mm}}cccccccccccc}
\hline
Field/CCD (field.ccd) & ID & RA (J2000) & Dec (J2000) & xpos (ccd) & ypos (ccd) & MAG & uncertainty & flag & COL & uncertainty & flag \\
\hline
1.02   &  410 &  22 52 47.929 & +62 30 57.08 &   613.729 &  2634.474 &    11.723  &    0.010 &  OO   &  0.763  & 0.014  & OO  \\
 1.04   & 10 &  22 57 13.885 & +62 41 35.55  &  774.439 &   119.826  &   11.596  &    0.010 &  OO   &  0.589 &     0.013  & OO   \\
\hline
\end{tabular}
\end{center}
\end{table*}

\subsection{Image Processing and Optimal Photometry}
\label{subsec:phot}

Image processing and data reduction was performed in a consistent manner for all datasets.
The individual frames were bias subtracted using a median stack of
several bias frames.  The frames were then flat fielded using twilight
sky flats taken on one of the photometric nights. Only sky flats with
peak counts of less than 30,000 were used, to avoid non-linearity
effects.  A bad pixel mask was constructed for all frames by flagging
all pixels which deviated by more than 10$\sigma$ from the median of a
ratio of two flat fields. In addition, a smoothed version of a long
exposure was subtracted from the long exposure, and all pixels with a
value lower than 1000 were masked.  These procedures accounted for
most bad pixels, but left a number of bad columns unmasked. These
columns were identified by hand and added to the bad pixel mask.

The processed CCD frames were then analysed using optimal photometry,
as implemented by the {\sc cluster}
software described in detail by \cite{naylor02}, with modifications 
described by \cite{littlefair05}. Advantages of this
approach over classical aperture photometry include better
signal-to-noise ratios, and robustly determined uncertainties for each
observation.  Photometric measurements which are deemed of poor quality are
flagged, rather than rejected. The reduction of our datasets closely follows that of
\cite{littlefair05}. Profile corrections (analogous to aperture corrections
in classical aperture photometry) were allowed to vary as a 2nd order
polynomial function of position upon the CCD.

After profile correction, the photometric measurements were adjusted
for any difference in the airmass and transparency for each frame, by
determining a relative transparency correction from the bright stars.
Before this process we added an additional, magnitude-independent error of 0.01 mags to the results 
of each frame, in order to yield a plot of $\chi^2_{\nu}$ versus signal-to-noise ratio
that was flat and had a modal value of approximately 1. In the absence of colour
information, transparency correction may introduce errors into our relative photometry because of the 
colour dependence of extinction. Following the analysis of
\cite{littlefair05}, we find that any errors introduced should be
smaller than 3mmags. Hence we are confident this effect is negligible for the purpose of our analysis. 
An astrometric solution was achieved through comparison with a 2MASS
catalogue of the same region. A 6-coefficient solution to 
yielded a rms discrepancy in positions which was always less than 0.14 arcseconds.

\subsection{Transformation to a standard system}

Our aim for this dataset was to obtain high standard relative
photometry of Cep OB3b. This meant that we did not measure colour
information for each star at each epoch, and hence we are unable to
tie our photometry to a standard system directly. Instead, we use the BVI
photometry of \cite{mayne07} throughout this paper.

\subsection{Final Dataset}
\label{subsec:data}

\begin{figure}
  \includegraphics[scale=0.35,angle=90,trim=80 80 0 50]{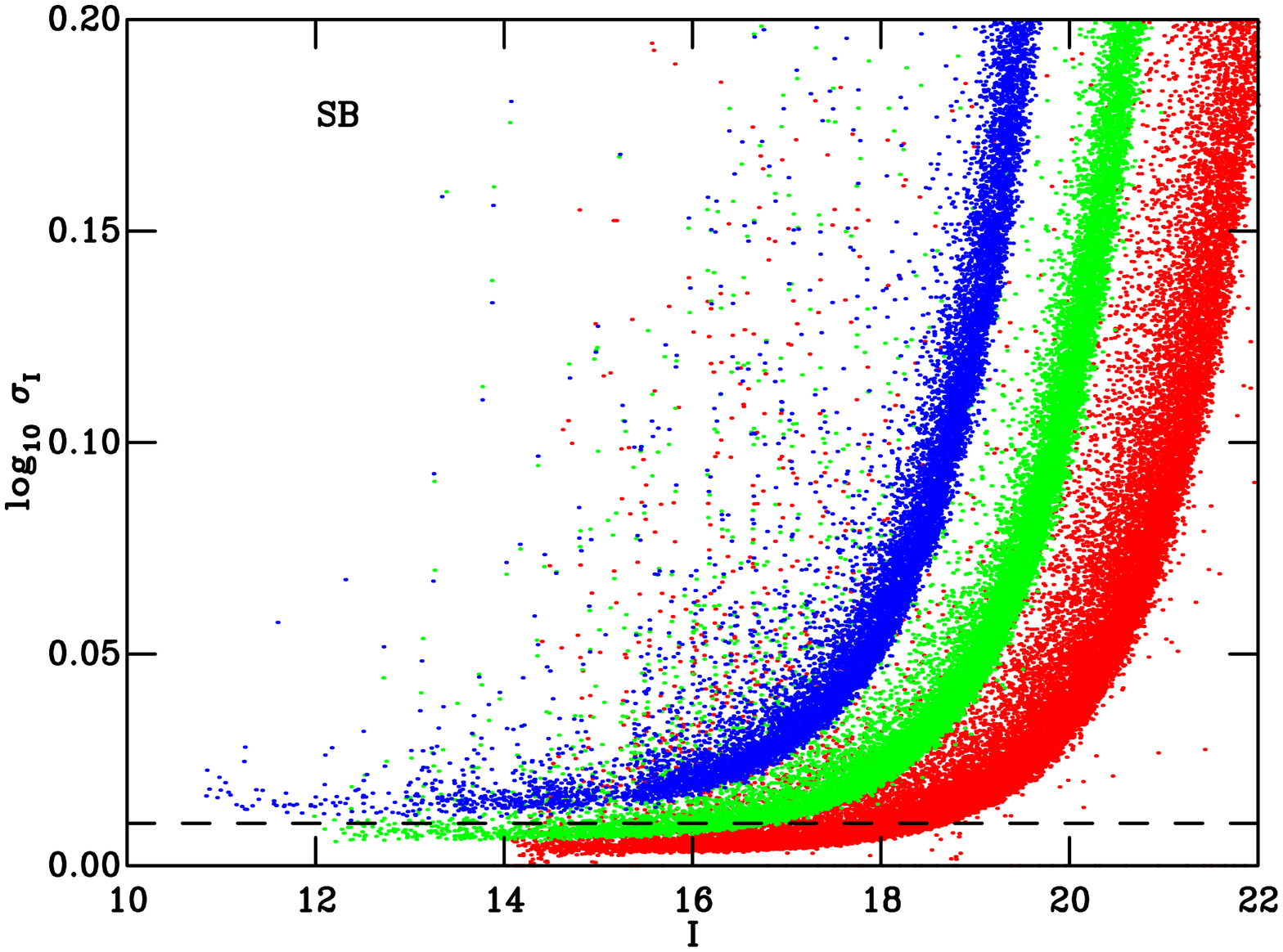} 
  \includegraphics[scale=0.35,angle=90,trim=80 80 0 50]{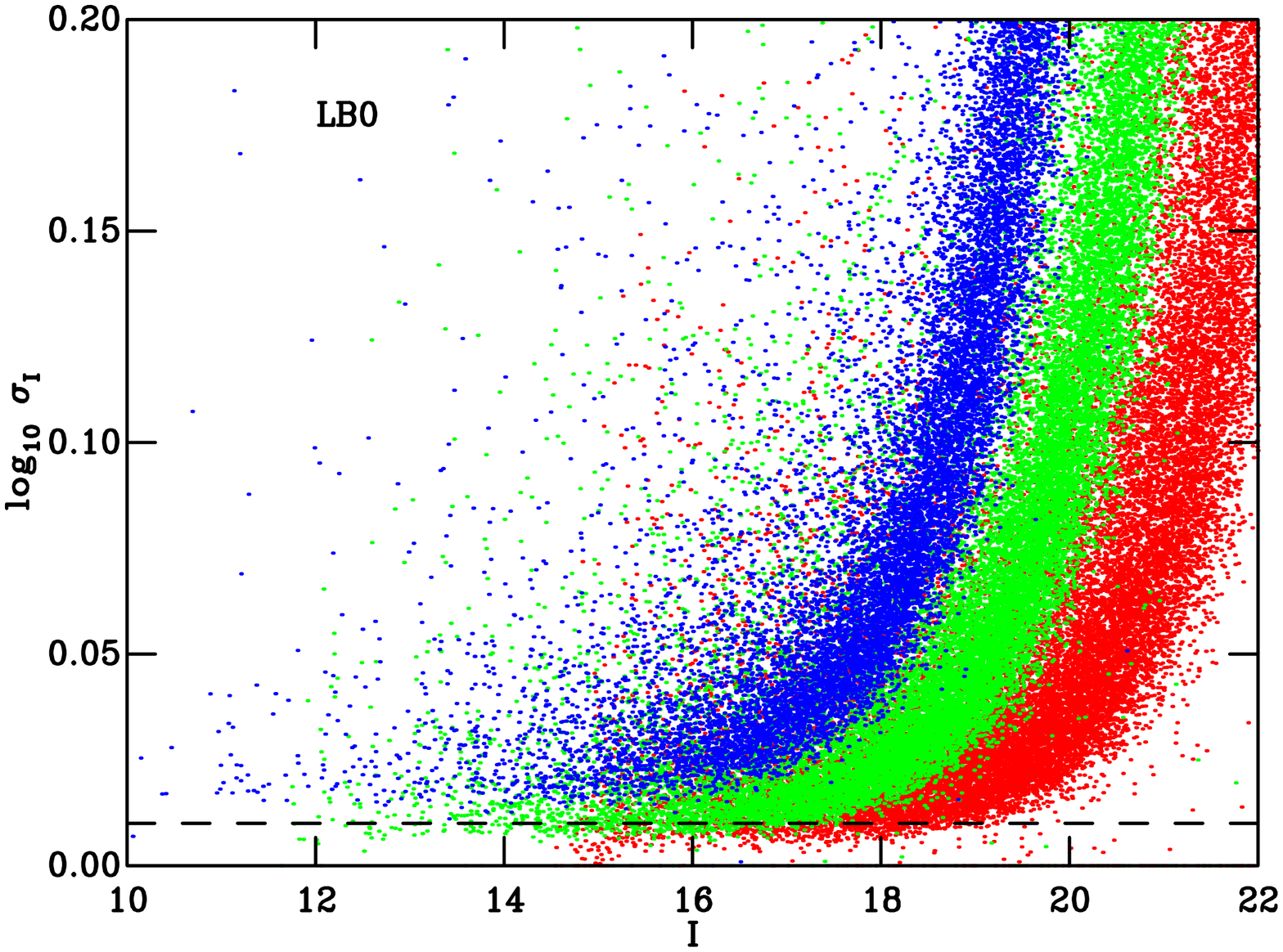}
  \includegraphics[scale=0.35,angle=90,trim=80 80 0 50]{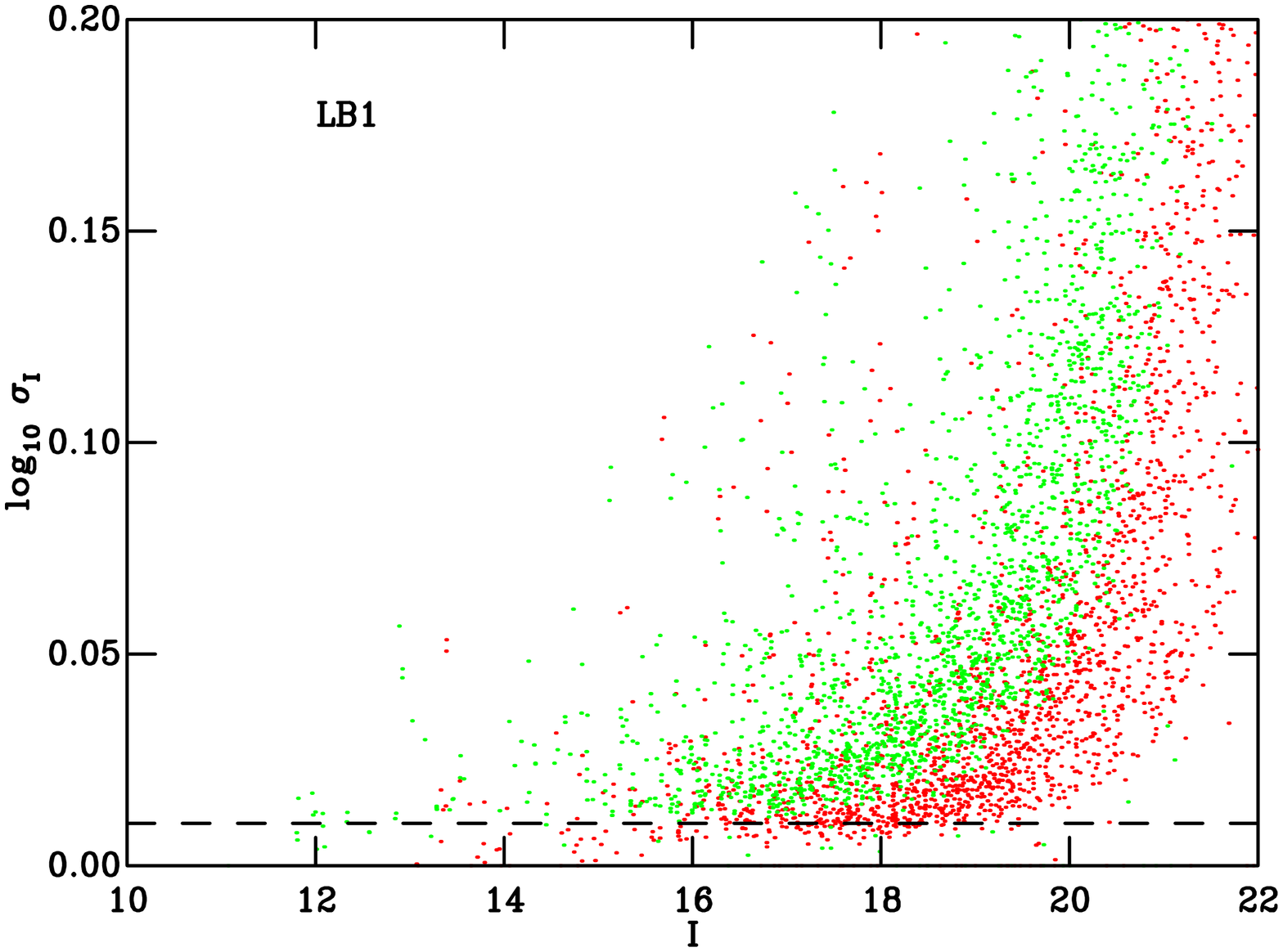}
\caption{Scatter in the photometry of stars as a function of
brightness in the SB (top), LB0 (middle) and LB1 (bottom) datasets. The 300s exposures are plotted 
in red (dark grey), the 30s exposures in green (light grey) and the 5s exposures in blue (black). The 
dotted line marks a photometric precision of 1\%}
\label{fig:rms}
\end{figure}

The final data set consists of lightcurves for 42962 stars.  Rather than
combine data with very different sizes of error bars, three
lightcurves were produced for each star, resulting from the 300, 30
and 5 second exposures (where available). This was done for each of the 
SB, LB0 and LB1 datasets, resulting in a maximum of 8 lightcurves for each star.
Figure~\ref{fig:rms} shows the RMS variability in magnitudes for our stars, plotted as a function of 
magnitude. This plot shows the high internal accuracy reached in
our dataset. In both the SB and LB0 datasets, obtained with the INT WFC, the 300s and 30s datasets 
have an internal accuracy of better than 1 per cent, whilst the 5s second dataset has an internal 
accuracy of 2 per cent. The gradual decrease in internal accuracy with decreasing
exposure time is most likely due to a corresponding decrease in the number of stars available to 
perform the transparency correction. The LB1 dataset, taken on the FTN, reaches an accuracy of 1 
per cent in both the 30 and 300s exposures.

\section{Detection of Periodic Variables}
\label{sec:detect}
To search for periodic variables within our datasets, Lomb-Scargle periodograms
were calculated independently for each of the SB, LB0 and LB1 datasets. Frequencies
corresponding to periods from 0.1--12 days were searched for the SB dataset, and
4--25 days for the LB0 and LB1 datasets. In each case the lower value is set by the "mean nyquist" 
frequency of the lightcurve.

Searching for periodicities in photometry of young stars is a challenging task. The unevenly spaced
data sampling means that analytical estimates of the false alarm probability (FAP)
 \citep[e.g.][]{horne86b,nemec85} are rendered invalid. Worse still, many young stars
show irregular variability as well as, or instead of, periodic photometric modulations from
surface spots. The interaction of this intrinsic variability with the often patchy sampling window
can quite easily introduce peaks into a periodogram which appear significant based on a
simple FAP cut.

The inadequacy of analytical estimates of FAP is often sidestepped by estimating the FAP
via a Monte-Carlo simulation; the FAP is set to be the fraction of simulated lightcurves where 
the peak power exceeds the observed peak power. The simulated lightcurves can either have 
pure Gaussian noise added, or some correlated noise model can be assumed. Eventually,
one ends up with a cut-off level for the peak power, above which a period is judged to be significant.

The goal of these Monte-Carlo simulations is to estimate the heights of spurious peaks which may arise in the
periodogram due to a combination of imperfect sampling, and variability in the lightcurve. However, they can only realistically account for variability in the lightcurve due to photometric error. For a large number of young stars, intrinsic variability is larger than the photometric errors. In addition, colour-dependent effects mean that transparency variations can be imperfectly corrected, introducing a night-to-night variability that Monte-Carlo simulations don't attempt to account for. Selecting periodic variables solely on the basis of Monte-Carlo simulations to asses FAP can thus lead to significant contamination of a dataset by spurious periods. The large number of objects with periods very close to 1d in the NGC 2264 dataset of \cite{lamm05} and \cite{lamm04} may be an example of this.

\begin{figure}
\begin{center}
\includegraphics[scale=0.37,angle=90,trim=90 0 0 80,clip]{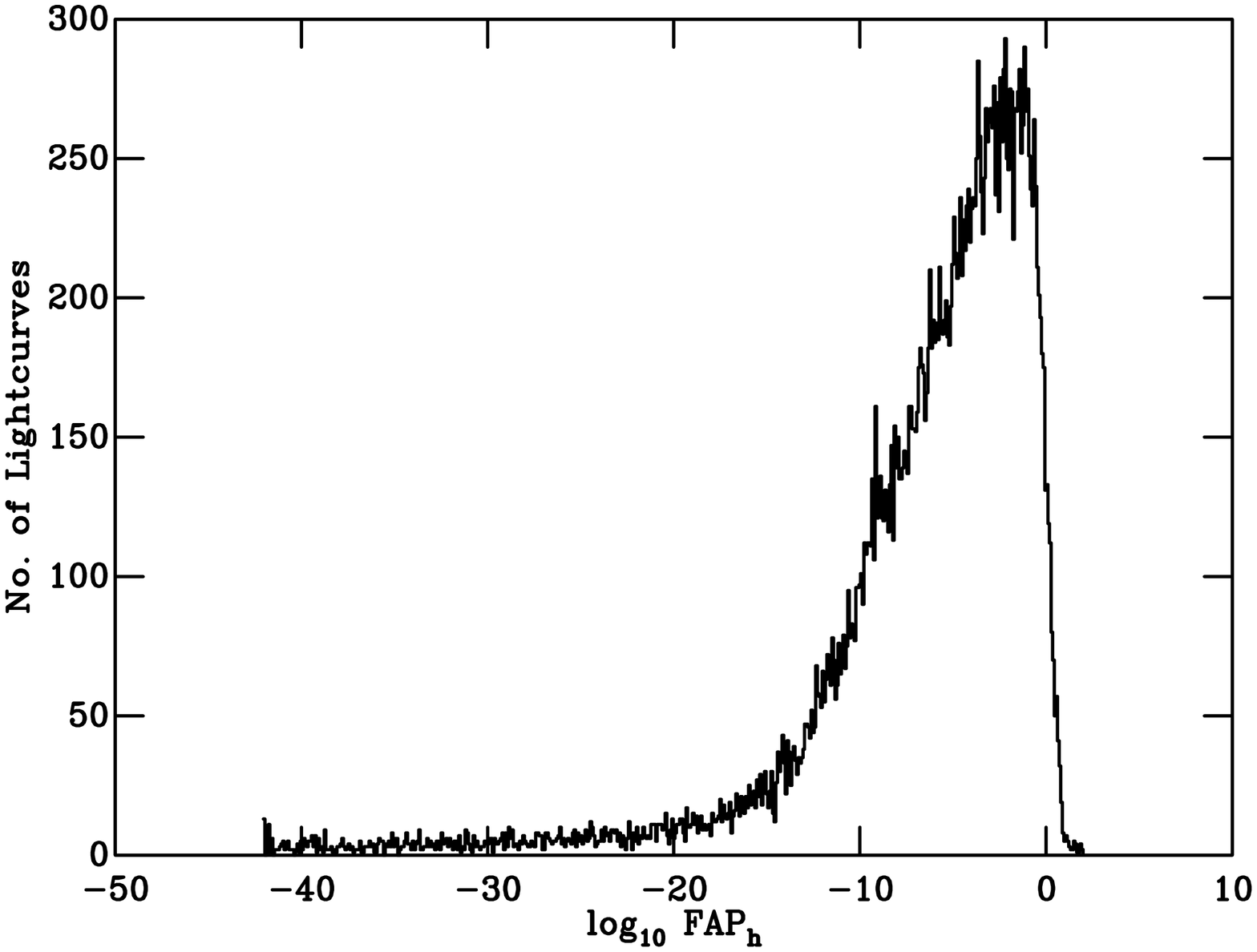} 
\includegraphics[scale=0.37,angle=90,trim=90 0 0 80,clip]{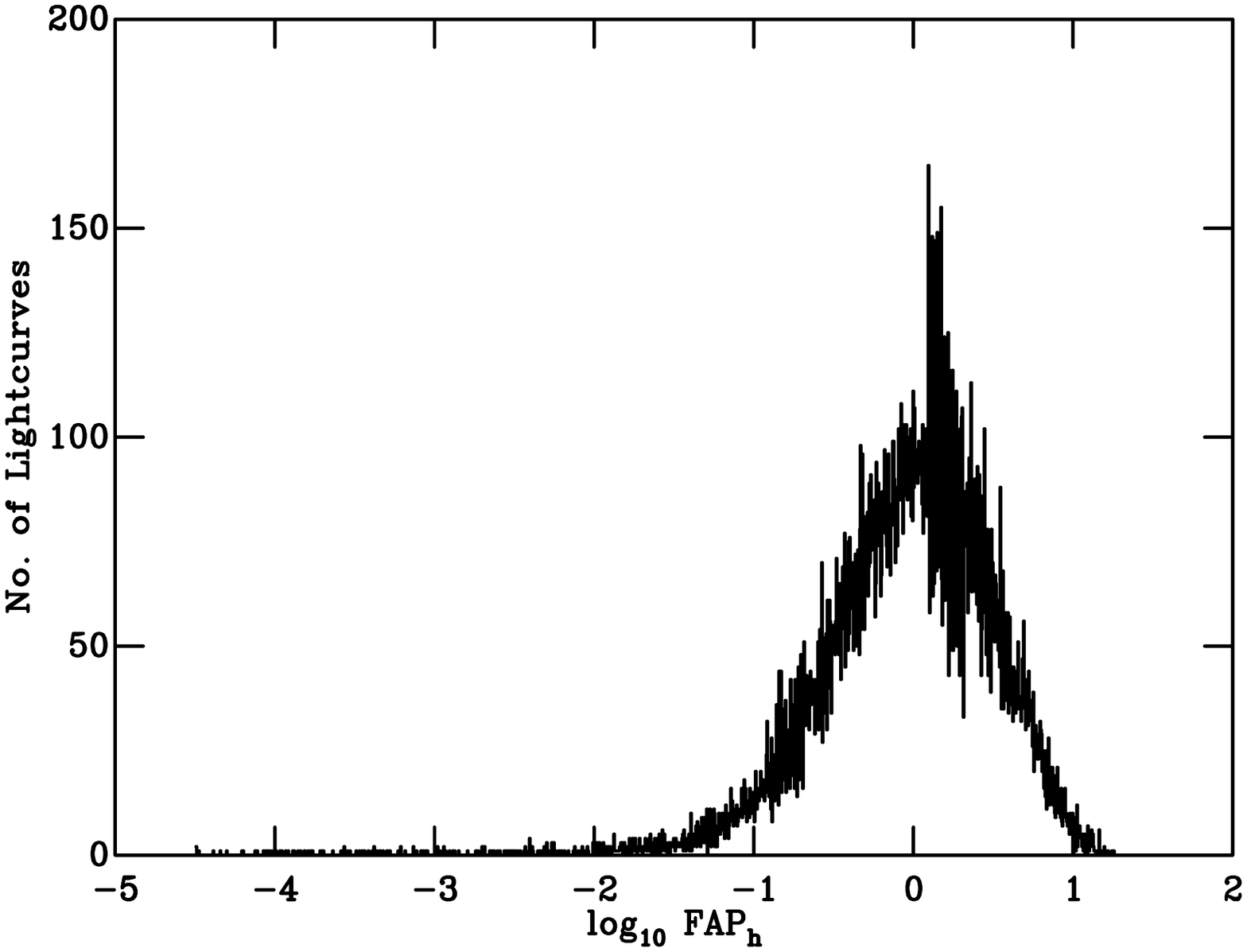} 
\includegraphics[scale=0.37,angle=90,trim=40 0 0 80,clip]{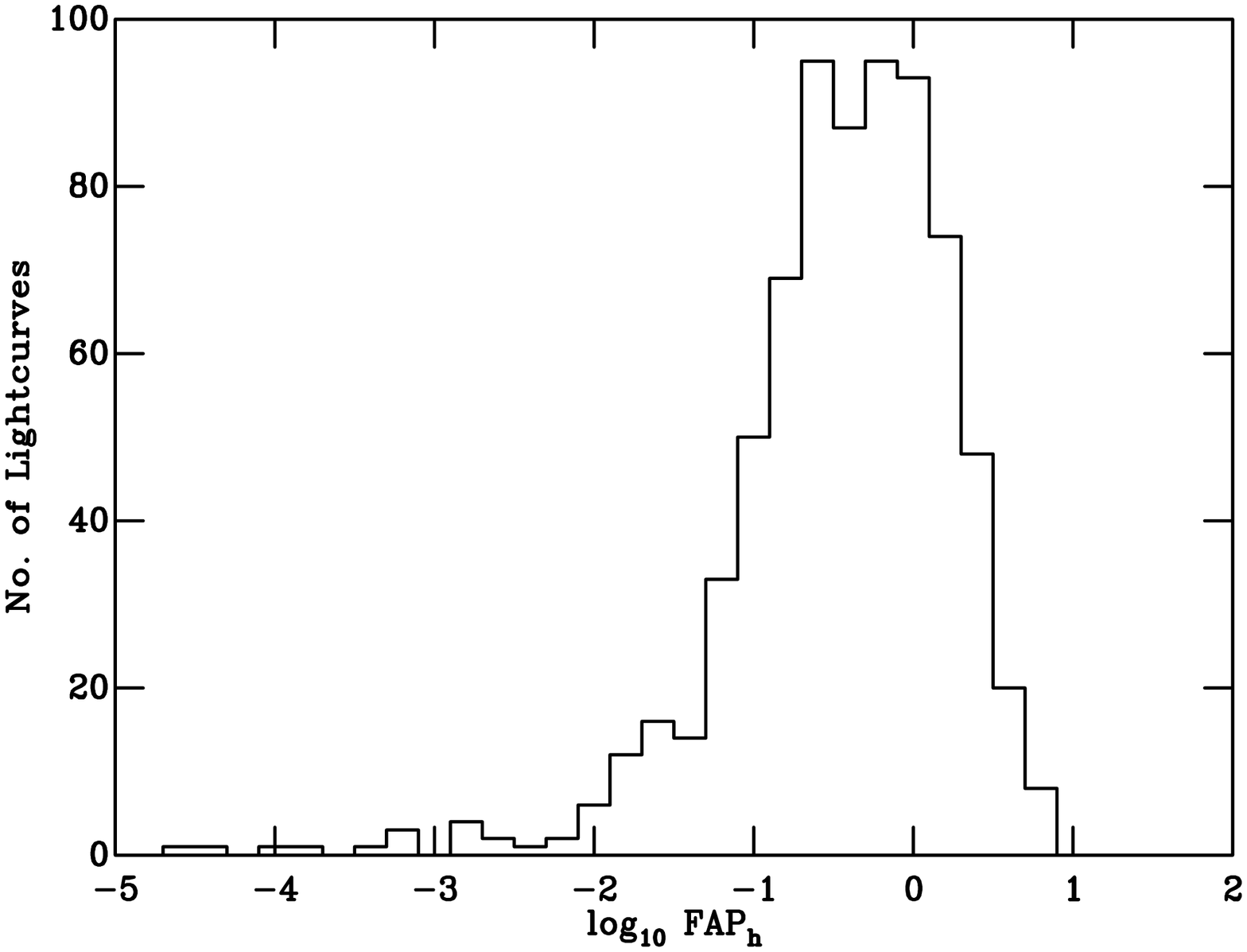} 
\caption{The distribution of FAP$_h$, for the SB (top), LB0 (middle) and LB1 (bottom) datasets.}
\label{fig:fap}
\end{center}
\end{figure}

What we are aiming for is a sample of genuine periodic variables, which is relatively free from contamination
by spurious periods. Since we have a large sample of lightcurves to search, one approach is to use the lightcurves themselves as a means of estimating the height of spurious peaks introduced into the periodogram by all
factors affecting the dataset, from imperfect photometry to intrinsic variability. We calculate an analytical
FAP, as estimated by \cite{horne86b}, FAP$_h$, for the strongest peak in all our lightcurves. Assuming that
most stars in our field of view are not periodic, the distribution of FAP$_h$ is determined by the interplay of stellar variability, finite signal/noise and systematic effects in our photometry with the sampling windows of our lightcurves.
Candidate periodic variables can be selected as outliers in the FAP$_h$ distribution. The value of FAP$_h$ to make a cut at is selected from a histogram of  the FAP$_h$ values from our lightcurves. The value of FAP$_h$ is chosen by-eye from the histograms shown in figure~\ref{fig:fap}. After applying the additional selection criteria described below, the level of contamination is assessed by visual inspection of randomly selected folded lightcurves, and the chosen FAP$_h$ value is adjusted to keep the contamination at an acceptable level.  During the  visual examination of folded lightcurves, most rejected candidates were those whose periodic nature rested on only a few, discrepant points (such as might occur from two flares repeated days apart, for example). Lightcurves were inspected by a single author (Littlefair), and the process repeated until the results were acceptable. We settled on a FAP$_h$ cutoff of $10^{-35}$ for the SB dataset, and a FAP$_h$ cutoff of 0.005 for the LB0 and LB1 datasets.

\begin{figure*}
\begin{center}
\includegraphics[scale=0.55,angle=-90,trim=0 0 0 0,clip]{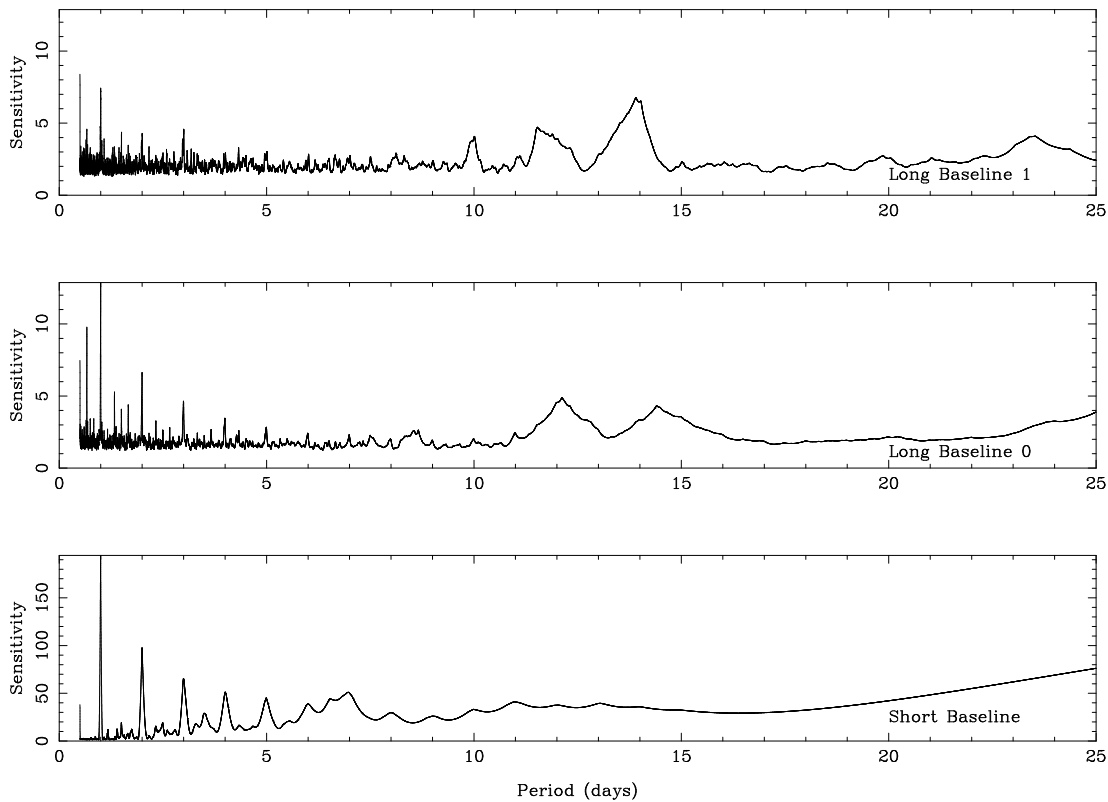} 
\caption{The Saunders statistic, as calculated from a representative lightcurve from each dataset, in 
which more than 60 per cent of the data points in the lightcurve were
unflagged.}
\label{fig:saunders1}
\end{center}
\end{figure*}

\begin{figure}
\begin{center}
\includegraphics[scale=0.6, angle=0,trim=0 0 0 0,clip]{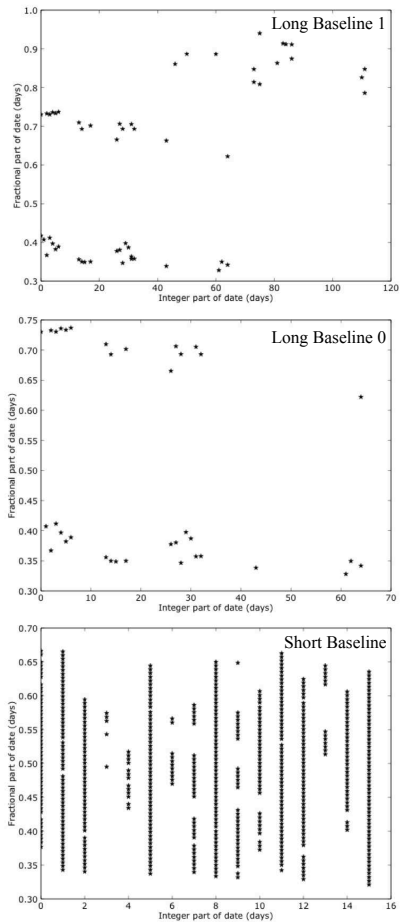}
\caption{Time sampling of our datasets. The x-axis shows the integer part of the time
elapsed since the first observation. The y-axis shows the fractional part of the time
elapsed since the first observation.}
\label{fig:coverage}
\end{center}
\end{figure}

A FAP cut taken alone will not reject some objects which should not be included in a sample of 
periodic variables. Two examples are very irregular variables and stars with poor rotational phase coverage.  
Repeating patterns in bright, irregular variables can add significant power to a Lomb-Scargle periodogram, and stars 
with poor phase coverage must necessarily be considered dubious periodic candidates. To remove these interlopers 
we applied two further cuts: one on the $\chi_{\nu}^2$ with respect to a sinusoidal fit at the suggested period,  and 
another on a statistic designed to maximise phase coverage, the S-statistic \citep{saunders06}. The S-statistic is the 
sum of the squares of the phase differences between adjacent data points, after the data has been sorted according 
to phase. It is normalised by division by the number of datapoints, such that uniform phase coverage gives an S-
statistic of unity, and the S-statistic rises with increasingly irregular phase coverage. The S-statistic is especially good 
at removing spurious periods arising from the 1 day$^{-1}$ natural frequency that is introduced into the data the 
rotation of the Earth, and observing at a single site. The S-statistic for our datasets as a function of period is shown in 
figure~\ref{fig:saunders1}. This figure shows the improvements in phase coverage offered by the long baseline datasets; particularly for periods around multiples of a day\footnote{The long period datasets show spikes in the Saunders statistic  at periods between 11 and 15 days. This means the phase coverage of these datasets degrades when the data is folded on these periods. In figure~\ref{fig:coverage} we show a graphical representation of the temporal coverage of each of our datasets. Looking at this figure, we can understand that the spikes in Saunders statistic at 11-15 days arise because the long period datasets have a tendency towards gaps in coverage lasting 5-8 days. We are not able to offer a definitive explanation of the reason for these gaps in coverage, but we can speculate that it is linked to the typical duration of bad weather at our observing sites.}. We calculate the S-statistic for all the periodic candidates in our sample, after converting to phase using the best fitting period as deduced from the Lomb-Scargle periodogram. We rejected all stars with an S-statistic greater than 30 (SB), 5 (LB0),  and 7 (LB1), or with periods in the ranges 0.9--1.05\,d, 0.495--0.505\,d and 0.325--0.335\,d.  The $\chi_{\nu}^2$ cut  was established via a by-eye inspection of 200 randomly selected candidate periodic variables. The value chosen rejected erratic variables, whilst preserving the majority of stars which were clearly periodic, but non-sinusoidal. We are thus confident that the $\chi_{\nu}^2$ cut is not rejecting periodic variables with a high signal/noise ratio whose lightcurves are not purely sinusoidal. All stars with $\chi_{\nu}^2>1.3$ were rejected. 

The selection process outlined above is far from perfect, but there is no ideal solution yet found to the problem of searching for periods in an imperfect dataset of objects which exhibit significant intrinsic variability. The selection criteria adopted above are conservative; the aim was not to discover all rotational periods, but to produce a catalogue of periodic stars which was relatively free from contamination by spurious periods.  To check our success in this aim we undertook a by-eye examination of 200 randomly selected folded lightcurves. This examination suggests that our final sample of periodic stars has a contamination by spurious periods of $\sim$3\%.

Once each dataset was searched for periods, we combined the results to produce our final catalogue of periodic 
variables. As each star could have a detected period in each of the datasets, we selected a period from the dataset 
which was most likely to be accurate. For periods longer than 7 days we adopted a period from the dataset with the 
longest baseline, whilst for shorter periods we adopted a period from the dataset with the densest time sampling. 
Where a dataset has lightcurves with differing exposure times we always used the deepest lightcurve in which a 
period was detected. In the vast majority of cases, periods from different datasets were consistent. Only 8 objects showed significant disagreement between datasets; these were long-period systems for which the SB dataset lacked sufficient baseline. The final periodic dataset consists of 709 periodic variables; 578 from the SB dataset, 88 from the LB0 dataset and 43 from the LB1 dataset. 704 of our periodic variables have BVI photometry from \cite{mayne07}. These 704 objects comprise our final catalogue of periodic variables, and are presented in Table 2, which is available in the online version of this paper.

\begin{table*}
\begin{center}
\caption[]{Data on the detected periodic variables is available in this format at http://www.astro.ex.ac.uk/people/timn/Catalogues/tables.html. This table is a sample only, to indicate the format of the online table. The flag column indicates data quality, and is explained in full at http://www.astro.ex.ac.uk/people/timn/Catalogues/format.html}
\begin{tabular}{@{\extracolsep{-1.25mm}}cccccccccccc}
\hline
Field/CCD (field.ccd) & ID & RA (J2000) & Dec (J2000) & xpos (ccd) & ypos (ccd) & Period & uncertainty & flag & MAG & uncertainty & flag \\
\hline
1.04  &   549  & 22 55 27.231 & +62 43 28.87  &  474.084  &  2343.941  &  2.089  &    0.010 &  OO  &  15.119  &    0.008 &  OO \\
1.04  &    43  & 22 57  2.094 & +62 44  7.04 &   319.608 &   374.983  &  2.630  &    0.010 & OO  & 15.297  &   0.008 &  OO  \\
\hline
\end{tabular}
\end{center}
\end{table*}

\subsection{Period completeness and selection effects}
\label{subsec:biasses}
\begin{figure}
\begin{center}
\includegraphics[scale=0.4,angle=0,trim=0 0 0 0,clip]{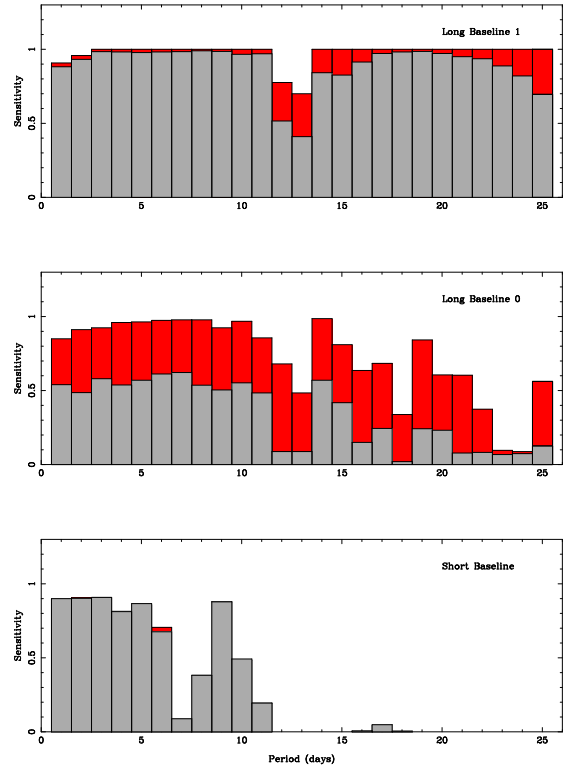} 
\caption{The fraction of correctly recovered periods, as a function of
period, for a representative lightcurve from each dataset. The red histogram shows the sensitivity to
sinusoidal periods with signal/noise ratios of 5, the grey histogram shows the sensitivity to sinusoidal 
periods with signal/noise ratios of 2.}
\label{fig:complete}
\end{center}
\end{figure}
\begin{figure}
\begin{center}
\includegraphics[scale=0.5,angle=0,trim=0 0 0 0,clip]{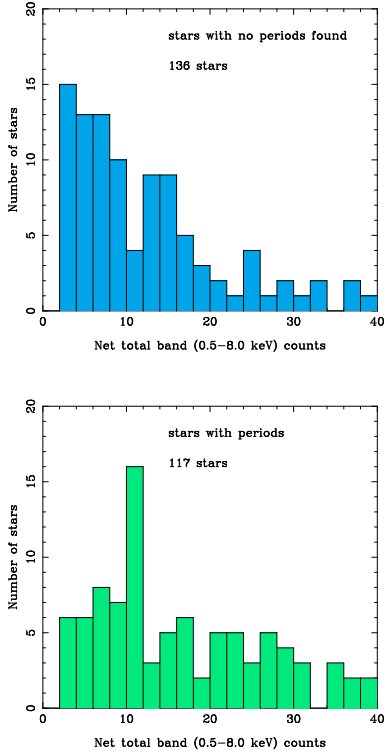} 
\caption{The histogram of X-Ray counts for objects with known periods
(bottom panel) and for objects with no known periods (top panel). The
X-Ray flux of objects for which we found periods is systematically
higher than for objects where no period was found.}
\label{fig:xray}
\end{center}
\end{figure}

We investigated our ability to recover a range of periods by introducing artificial sinusoidal signals 
into our data and checking to see if the correct period was recovered. Lightcurves from our datasets were selected at random, normalised and had an artificial sinusoidal signal added with a signal-noise of 2. For our dataset, this is equivalent to adding a 0.02 mag signal to an object with I=18 (the faintest magnitude for which report periods, and corresponding roughly to a mass of 0.2 M$_{\odot}$). Our simulation is thus directly comparable to the simulation by \cite{irwin08} for their periodic dataset in NGC 2362. We also performed a simulation in which sinusoidal signals with a signal-noise ratio of 5 were added to a normalised lightcurve.

The period selection process described above was then performed on our artificial 
lightcurves. A period was deemed correctly recovered if the if the recovered period was within 10 per 
cent of the injected period. Figure~\ref{fig:complete} shows the results. The SB dataset finds 
essentially all periods below 7 days, but is incomplete above that. This is expected from the relatively short baseline 
of this dataset; indeed the reason so few periods are recovered above 10 days is that these periods fail our test based on the Saunders statistic.  The LB0 dataset improves upon this substantially, but is still not complete above 10 days. 
Because of the relatively sparse sampling of this dataset, completeness for the LB0 dataset is a strong function of the injected signal size. For weak signals the completeness never rises much above 50 per cent, whereas this dataset recovers most stronger periodic signals below 10 days. The LB1  dataset is largely complete at all periods, with the exception of a small drop in completeness between 12--14 days, however, the small spatial coverage of this dataset means that it will not be available for many stars.

This simulation shows that all datasets are sensitive to short periods (below 7 days). In practise, the 
SB dataset will be much better at recovering period shorter than 7 days. This is because the 
simulation described above does not take account of the stochastic variability often seen in young 
stars. This variability can be considered red noise, and has characteristic timescales of days to 
weeks. As discussed by \cite{littlefair05}, dense sampling of lightcurves makes it much easier to 
recover periods in the face of this red noise. Therefore, we prefer periods from the SB dataset for 
periods less than 7 days.

In figure~\ref{fig:xray} we show the X-ray counts from \cite{getman06} for those objects for which we 
found periods, and for those objects in which no periods were found. Objects with detections in the X-ray catalogue of \cite{getman06} were matched with our optical catalogue using a 3 arcsecond matching radius.  A cut in colour-magnitude space (see section~\ref{subsec:cleanup}) was applied to our optical catalogue to reduce the contamination from foreground objects. The matches were then divided into two groups; those in which we detected periods, and those in which no period was found. We see a suggestion that the stars in which we found periods are significantly more X-ray luminous than those stars for which a period was not found. This result is not formally significant; A 1-D K-S test gives a 13 per  cent chance that they were drawn from the same parent distribution. Thus it is possible, though not certain, that our catalogue of periodic objects is biassed against X-ray faint sources. This bias is presumably introduced because strong starspot activity is associated with increased X-ray emission. Whether this introduces bias into our  period distribution remains to be seen. There is no evidence for a correlation between X-ray counts and rotational period {\em within} the stars for which periods were found. 

To summarise; our survey is sensitive even to weak periodic signals below 7 days and down to masses of 0.2 M$_{\odot}$. For longer periods, completeness is somewhat lower, and we are, generally speaking, only sensitive to stronger periodic signals (S/N $\ge$ 5). Thus, our periodic dataset is biassed to some degree towards shorter rotational periods.

\subsection{Rejection of background and foreground objects}
\label{subsec:cleanup}
\begin{figure}
\begin{center}
\includegraphics[scale=0.4,angle=90,trim=30 0 0 80,clip]{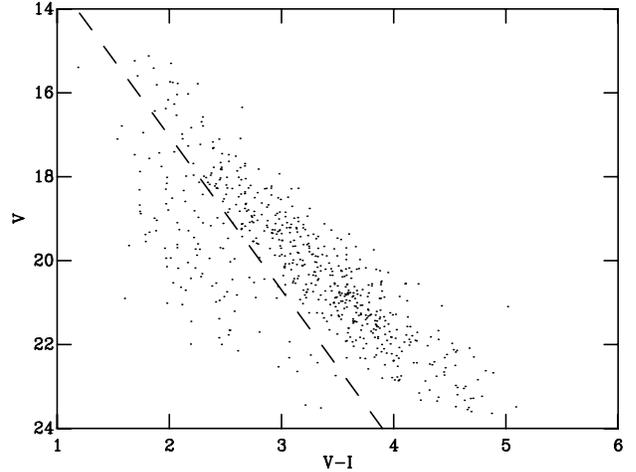} 
\caption{The colour magnitude diagram of periodic variables in Cep
OB3b. An initial colour cut, used to select PMS objects is plotted
with a dashed line.}
\label{fig:cmd}
\end{center}
\end{figure}

\begin{figure}
\begin{center}
\includegraphics[scale=0.4,angle=90,trim=30 0 0 80,clip]{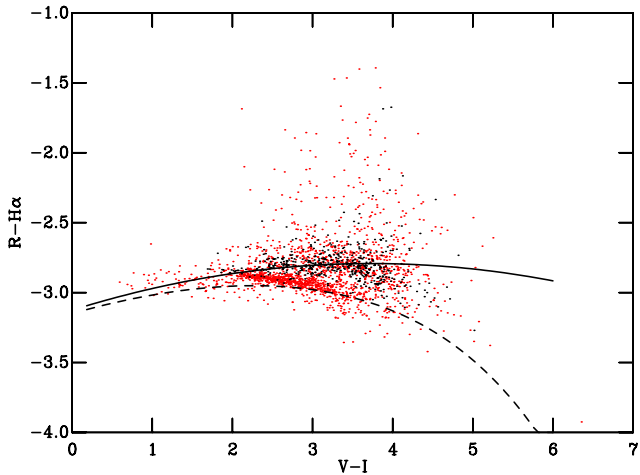} 
\caption{The $(R-H\alpha)$ vs. $(V-I)$ colour-colour diagram for Cep OB3b. The
red (light grey) points represent all the stars for which we have photometry. The
black dots are those periodic stars which survived the cut in the $V$ vs. $V-I$
cmd. The pre-main-sequence locus is plotted as a solid line, and the locus used to
reject background giants is plotted as a dotted line.}
\label{fig:col}
\end{center}
\end{figure}

There will also be some contamination of our sample by non-PMS stars. Whilst variability is a very good indicator of 
youth, periodic variables do exist amongst the field population, and these should be removed from our sample. The 
most practical way of removing foreground and background objects is via cuts in colour-magnitude space.  PMS 
objects are larger and correspondingly brighter than main sequence stars of a given colour and distance. A simple cut in colour-magnitude space can remove a large amount of the contamination by main sequence and other foreground objects. We decided on an appropriate cut by-eye, which is shown in figure~\ref{fig:cmd}. Of the 704 periodic stars, 553 have unflagged colours placing them above and to the right of the cut in figure~\ref{fig:cmd}. Our dataset is still not free of possible contamination. \cite{lamm04} points out that significant contamination from variable background giants (e.g. RR Lyrae stars) can lie within the PMS region as defined by the cut above, and a further cut is required to remove these objects. \cite{lamm04} suggest that the difference in $(R-H\alpha)$ colours between giants and main-sequence or pre-main-sequence stars allows for rejection based upon this colour index. In addition, the extra reddening experience by background giants gives them larger
$(V-I)$ colours than main-sequence or pre-main-sequence stars. Therefore, following \cite{lamm04}, we use the $(R-
H\alpha)$ vs. $(V-I)$ colour-colour diagram, shown in figure~\ref{fig:col}, to reject background giants. The locus of 
PMS stars was defined as the median $(R-H\alpha)$ colour as a function of $(V-I)$ for all periodic stars which 
survived our initial cut in $V$ vs. $(V-I)$. A quadratic fit to the median defines the PMS locus as
\begin{equation}
(R-H\alpha)_{locus} = -0.024(V-I)^2 + 0.179(V-I) - 3.126
\label{eq:locus}
\end{equation}
To reject background giants we define a lower cut-off, shown in figure~\ref{fig:col} as a dashed line, by $(R-H
\alpha)-1.65\delta$, where $\delta$ is a fit to the standard deviation in $(R-H\alpha)$ as a function of $(V-I)$, given by
\begin{equation}
\delta = \exp \left[ 0.66(V-I) - 4.23 \right].
\end{equation}

Of the 553 stars which passed our $V$ vs. $V-I$ cut, 481 had unflagged colours in both $(R-H\alpha)$ and $(V-I)$,  6 
of which were rejected as background giants. Our final dataset thus consists of 475 periodic variables which are very 
likely pre-main-sequence members of the Cepheus OB3b star-forming region.

\section{Age and Distance of Cep OB3b}
\label{sec:distance}

In \cite{mayne07} we derived ages for a group of well 
studied star-forming regions, including Cep OB3b, based on the fact that the 
luminosity of the pre-main-sequence declines with age.
Thus the derived age clearly depends on the distance assumed for each group, 
which in \cite{mayne07} we took from the literature.
Realising these literature distances were the major source of uncertainty in the
ages, \cite{2008MNRAS.386..261M} derived consistent distances with robust
uncertainties by fitting the main-sequence stars in each region using the 
$\tau^2$ method of \cite{2006MNRAS.373.1251N}.
We have applied the same methods to the $\gamma$ Vel association and Vel OB2 
 \citep{jeffries09} and NCC2169 \citep{2007MNRAS.376..580J}, which along
with the regions presented in \cite{2008MNRAS.386..261M} represent a 
self-consistent set of ages and distances for a significant group of young 
clusters and star-forming regions.

Clearly we must attempt to fit Cep OB3b into this system, but
although we attempted this in 
\cite{2008MNRAS.386..261M}, we failed to do so because we had only a 
few stars which could be 
used for main-sequence fitting, and they gave an answer sharply at variance
with the literature values.
In what follows we use the stars from Cep OB3a as well as Cep OB3b, and 
still obtain an answer which
is different from that determined by others, but we now understand that the
difference is due to the earlier work using a \cite{Schmidt-Kaler} 
main-sequence (as opposed to his zero-age main-sequence), 
which is significantly brighter than more modern work.

To determine an age in the same way as \cite{mayne07} we must 
place the pre-main-sequence of Cep OB3b in a $V_0$ vs $(V-I)_0$ CMD, and compare
it with similar sequences for other star-forming regions.
If we carry out such a comparison by plotting the members of each group in 
a CMD the number and spread of the data points make the plot
impossible to interpret.
Instead, as described in \cite{mayne07}, we use splines 
fitted through the members.
To place Cep OB3b in such a diagram we need the apparent distance 
modulus (to correct $V$), and $E(V-I)$ to correct $V-I$.
In what follows we derive these by dereddening and then fitting the 
main sequence, in an identical fashion to \cite{2008MNRAS.386..261M}.

\subsection{MS star Individual reddenings and extinctions}

\begin{figure}
\vspace{70mm}
\includegraphics{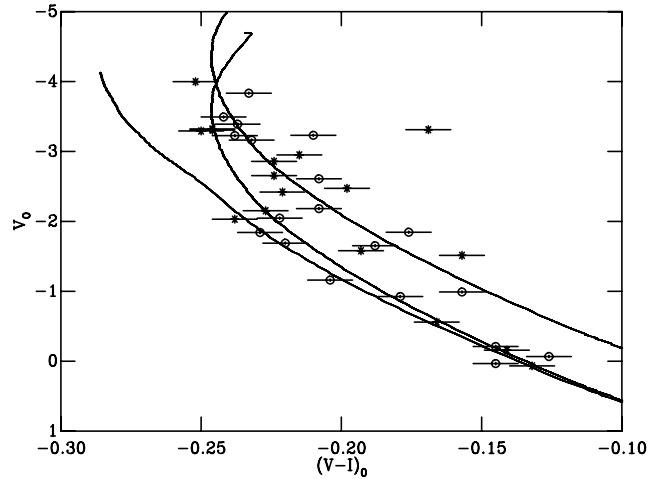}
\caption{
The members of Cep OB3a (circles with error bars) and Cep OB3b (stars with
error bars).
The photometry \citep[from][]{1959ApJ...130...69B}, and the colours
have then been de-reddened and the magnitude corrected for extinction by
deriving colour excesses in the $U-B$ vs $B-V$ plane.
The magnitudes have the been converted into absolute magnitudes using the
best-fitting distance modulus of 8.8.
The curves are Geneva-Bessell isochrones for (from faintest to brightest) 
3Myr, 10Myr and a 3Myr equal-mass binary sequence.
}
\label{ms_cmd}
\end{figure}

\cite{pozzo01} provided a list of members of both sub-groups 
with photometry in \cite[][hereafter BHJ]{1959ApJ...130...69B}.  
\footnote{The star numbers in BHJ for these are as follows.
Sub-group a: 14, 18, 19, 23, 37, 44, 46, 47, 50, 54, 56, 59, 
66, 68, 69, 70, 75, 76, 77.
Sub-group b: 2, 8 , 10, 11, 15, 16, 17, 20, 22, 24, 25, 26, 31, 33, 39, 40, 41}
We dereddened each of these in $U-B$ vs $B-V$ and used the implied 
extinction in $V$ to create the CMD shown in Figure \ref{ms_cmd}.
To do this we have used the extinction vectors of \cite{1998A&A...333..231B}, 
and a Geneva-Bessell main-sequence \citep[see][]{2001A&A...366..538L,
1998A&A...333..231B, 2008MNRAS.386..261M}. The mean extinction of the 
sample is $E(B-V)=0.79$ with an RMS of 0.16 mags.

\subsection{True distance modulus}
\label{true_dist_mod}

For our distance modulus determination we must ensure we fit only those stars
which are on the main sequence
\citep[see the discussion in][]{2008MNRAS.386..261M}.
The bright limit to the main sequence is defined by the turn-off, which 
becomes fainter with age.
In Figure \ref{ms_cmd} we show the 3 and 10Myr isochrones, which make it 
clear that the turn-off is not a sharp transition.
We chose these ages as 3Myr is the youngest age available in the Geneva
isochrones, whilst it is unlikely that any members of Cep OB3 are older 
than 10 Myr.  
We therefore fit only those stars lying below $V_o=-1.5$, since brighter than
this the two sequences separate by more than 0.009 mags in $B-V$, corresponding
to half the uncertainty for the photometry quoted by \cite{1959ApJ...130...69B}.
The faint end of the main sequence, within the CMD, is approximately defined by the radiative-convective (RC)
gap, which is only defined (and observable) for clusters younger than 20 Myrs old  \citep{mayne07}. Although 
some stars blueward of the RC gap will not have reached the main-sequence, their position in CMD space
is very close to the main-sequence locus at these ages (see discussion in \citealt{2008MNRAS.386..261M}).
Even at the age of the Orion Nebula Cluster (2Myr) the stars bluer than 
$B-V=0$ are on the main sequence, so the entire remaining sample can be fitted.

\begin{figure}
\vspace{100mm}
\includegraphics{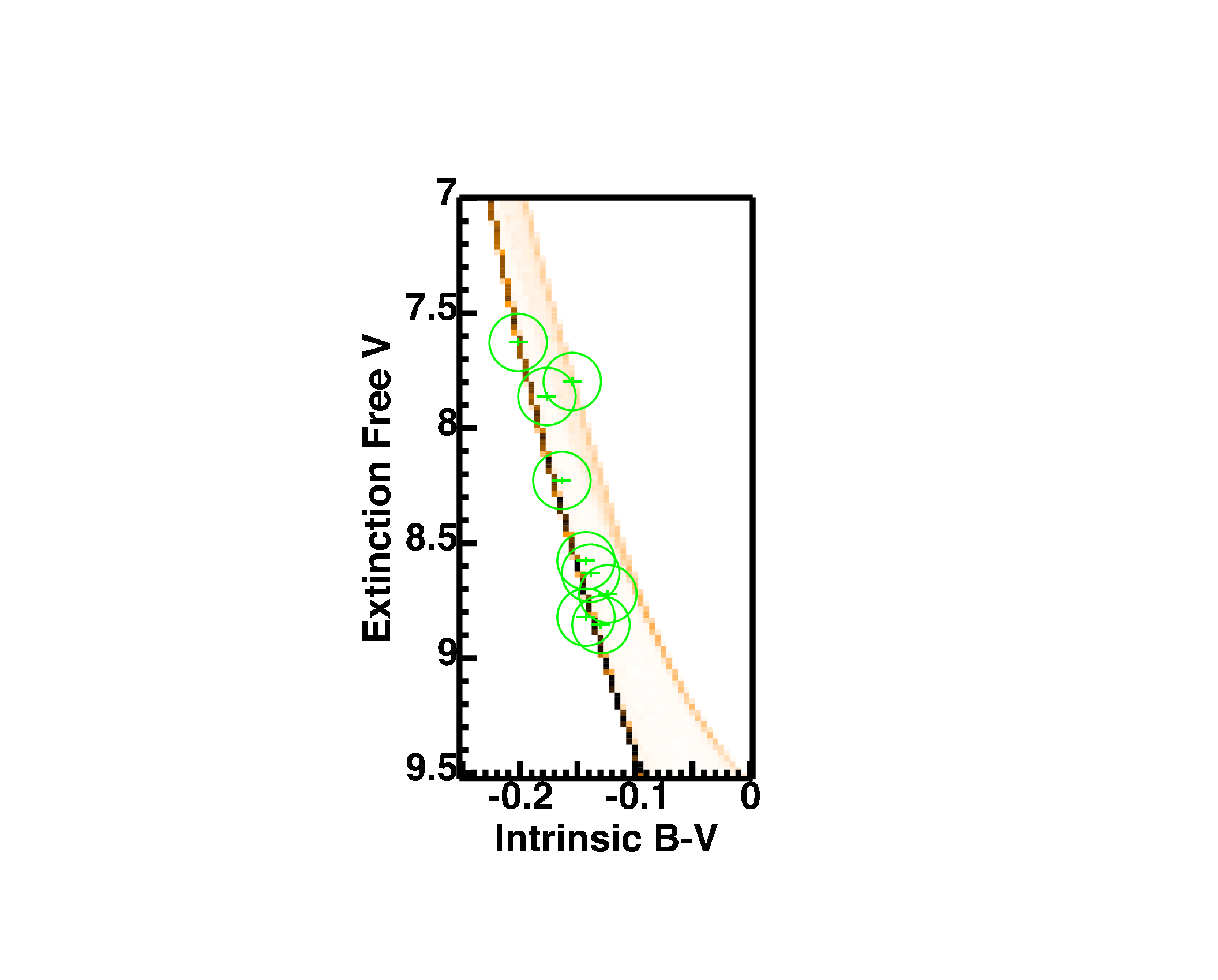}
\caption{
The fit to the main-sequence data.
The data are shown as circled error bars, and 
the colour scale is the best fitting model 
\citep[$\rho$ in Equation 3 of][]{2006MNRAS.373.1251N}.
}
\label{subset}
\end{figure}

We can now fit a main-sequence isochrone to the remaining data in Figure \ref{ms_cmd}, using the  $\tau^2$ fitting procedure described  in \cite{2008MNRAS.386..261M}, with the improvements in \cite{naylor09}. Using the Geneva-Bessell isochrones we obtain the fit shown in Figure~\ref{subset}, and a true distance modulus of $8.8 \pm 0.2$mags. 

\subsection{Apparent $V$-band distance modulus}

Since our aim is to correct the PMS data of \cite{mayne07} into 
the $V_0$ vs $(V-I)_0$ CMD, we need the apparent distance modulus for the
field from which those data were taken.
This is the same INT field as observed here, within which there are 
twelve BHJ stars, which give a mean 
$E(B-V)=0.83\pm0.11$ where the quoted uncertainty is the RMS about the mean.
We can then calculate $A_V$ as $3.26\times0.83=2.71$, and the apparent 
$V$-band distance modulus as $8.8+2.7=11.5\pm0.2$.
This is independent of any assumptions about $R = \frac{A_V}{E(B-V)}$, since changing $R$ changes
the extinction-free magnitudes of Figure \ref{ms_cmd}, and hence the
true distance modulus by an equal an opposite amount to the change in $V$-band
extinction.

\subsection{Reddening in $V-I$}
\label{reddening}

Finally we need $E(V-I)$ to correct $V-I$.
The usual way of proceeding at this point would be to use the ratio 
$E(V-I)/E(B-V)$ to calculate $E(V-I)$ from $E(B-V)$.
But in this case there is a subtle issue which must be addressed.
For a given column of interstellar material between the observer and a star 
the extinction (e.g. $A_v$) and reddenings (e.g. $E(B-V)$) depend on the 
colour of the star being observed.  
Therefore to specify the column density precisely, one must specify the colour
of the star for  which the measurement was made.  
For B-stars let us refer to $E(B-V)_B$, and for the red PMS stars $E(B-V)_R$.  
To construct the normal colour dependent reddening vectors 
one constructs a continuous sequence of ratios from say $E(V-I)_B/E(B-V)_B$ 
to $E(V-I)_R/E(B-V)_R$.  
Our case is somewhat different, we have measured $E(B-V)_B$ but
wish to know $E(V-I)_R$, and therefore need the ratio $E(V-I)_R/E(B-V)_B$.  
This can be obtained folding a blue stellar atmosphere through the 
appropriate filter responses, and then applying a given column density to 
the flux, to obtain $E(B-V)_B$.  
Using the same column density one can also obtain $E(V-I)_R$.  
Whilst \cite{1998A&A...333..231B} find that $E(V-I)_R/E(B-V)_R=1.44$ for 
$(V-I)_0=2$, we find that with identical atmospheres and the extinction
law of \cite{1989ApJ...345..245C} $E(V-I)_R/E(B-V)_B=1.26$ for red
stars of $(V-I)_0$ of approximately two.

In principle $A_V$ should show a similar effect, with red stars being less
affected by a given column density than blue ones, and so, in principle we 
have overestimated the $V$-band extinction for the red stars.
However, we have calculated this effect, and find it is less than 0.03 mags,
and so ignore it.
Thus our final estimate of $E(V-I)$ for appropriate stars of PMS colours is 
1.26$\times$0.83=1.05.

\subsection{Age estimate}
\label{age}
\begin{figure}
\vspace{70mm}
\includegraphics{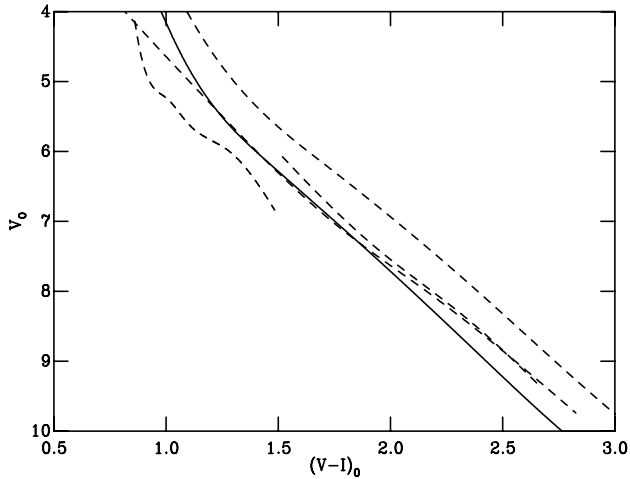}
\caption{
A comparison of the position of the PMS in a CMD of Cep OB3b with other
young groups.
The solid curve is for Cep OB3b.
The dashed curves are the sequences for (from left to right) 
the Orion Nebula 
Cluster (2Myr), NGC2362 (lower, 4-5Myr), IC348 (upper 4-5Myr) and h and 
$\chi$ Per (13Myr).}
\label{pms_cmd}
\end{figure}

Now we have distance moduli we can obtain a relative age for Cep OB3b
by comparing the absolute magnitude of its pre-main-sequence with those 
of other young groups.  Within any young ($<$5Myr) OB association or cluster, 
there is a large scatter in the luminosity of the pre-main-sequence members 
at any given colour \citep[for a discussion of this see, for example,][and references therein]{burningham05}.  
To make the comparison between groups easier, therefore, we
fit a curve through the members of each group, and then compare the positions of these fits
in an absolute-magnitude intrinsic-colour diagram.  The fits are taken from \cite{mayne07} and were constructed by 
binning the data by magnitude, finding the median
magnitude and colour within each bin, and then fitting a spline through the resulting
points.  The corrections to absolute magnitude and intrinsic colour were taken from
either \cite{mayne07} or \cite{2008MNRAS.386..261M}, except for Cep OB3b for which we used the
values derived above.  The result is shown in Figure~\ref{pms_cmd}, which shows the position 
of Cep OB3b very closely matches that of IC348 and NGC2362, both which were assigned 
ages of 4-5Myr,  and so we will use an age of 4.5Myr for Cep OB3b throughout this paper

\subsection{Comparison with literature values} 

In \cite{mayne07} we adopted an age of about 3Myr, and a distance
modulus of $9.65\pm0.2$ mags.
Our revised true distance modulus of 8.8 mags is much lower, yet our
new age is only slightly older.
The \cite{mayne07} distance was essentially the one derived by  
\cite{1993A&A...273..619M}, which relies on the \cite{Schmidt-Kaler} 
main sequence for stars around the solar neighbourhood.
This is about 0.8 mags brighter in $M_V$ than the Geneva-Bessell isochrone for 
3Myr, and indeed the {\em zero-age} \cite{Schmidt-Kaler} main sequence.
This effect alone would push Cep OB3b to an age of around 10Myr, but is
partially compensated for by our value of $E(V-I)/E(B-V)$.
The difference between the value we use, and say that of 
\cite{1998A&A...333..231B}, results in shift of 0.15mags in $V-I$, 
decreasing the age from 10Myr to the 4.5Myr adopted here.

Given the significant change in the distance compared with literature values,
and the low value of the $V-I$ colour excess we have used, we have checked
that the resulting age is consistent with other available age indicators. 
These are as follows.
\hfil\break
(i) Cep OB3b has a large R-C gap \citep[see][]{mayne07} 
suggesting an age of around 3Myr.
\hfil\break
(ii) The PMS shows a large ``age spread'' in the CMD, which
\cite{mayne07} show is characteristic of groups younger than about
5Myr.
\hfil\break
(iii) The turn-off age derived by \cite{1996A&A...312..499J} is 5.5Myr, and
although revised distance modulus would make this slightly younger, the
turn-off and PMS ages are clearly not grossly discrepant.
\hfil\break
(iv) There is still molecular material around the group (it is 
obvious in the optical images and the extinction is patchy), which is 
again normally associated with young groups.
Thus we conclude that all current age estimates are in concordance.

\section{Period distribution}
\label{sec:results}

\subsection{Mass dependence of rotation}
\label{subsec:mass}
The dependence of rotation period upon mass was first shown for the ONC by \cite{herbst02}, and later confirmed in 
NGC 2264 \citep{lamm04,lamm05} and IC 348 \citep{littlefair05}.  In general, lower mass stars rotate significantly 
faster than higher mass stars, both in the PMS phase and in young clusters \cite[e.g.][]{herbst07,scholz05,scholz09}. The reasons for this are not yet clear, though hints are emerging that it is related to a change in magnetic structure with mass \citep{scholz09,donati09}.

\begin{figure}
\begin{center}
\includegraphics[scale=0.6,angle=0,trim=120 20 100 0,clip]{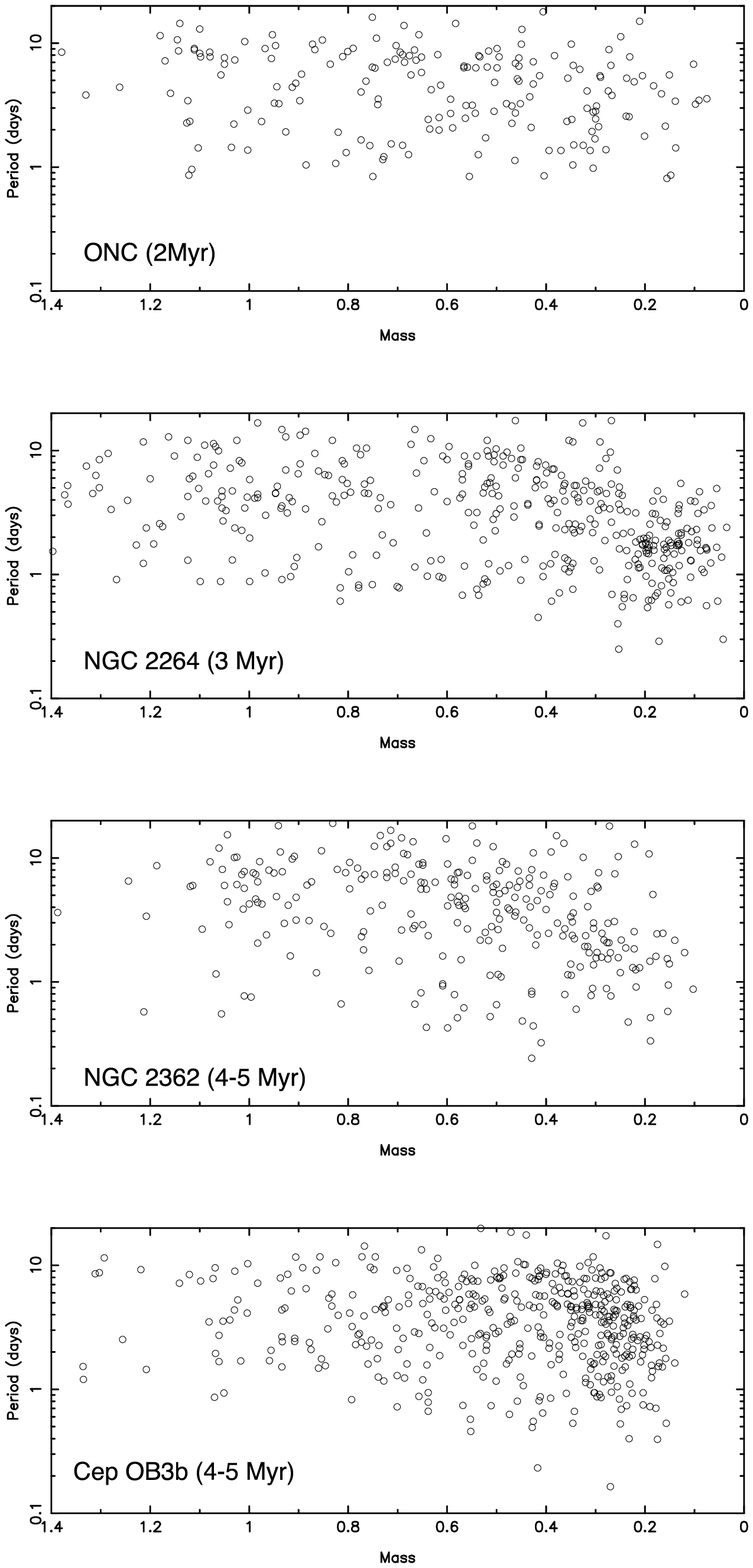} 
\caption{Rotation period plotted against mass for the stars in Cepheus OB3b, as well as for other clusters with large 
collections of rotation dat within the literature. Rotation periods were taken from \protect\cite{herbst02} (ONC), \protect
\cite{lamm04,lamm05} (NGC 2264) and \protect\cite{irwin08} (NGC 2362). See text for details of how masses were 
assigned.} 
\label{fig:pdist_v_mass_all}
\end{center}
\end{figure}

We calculated masses for our stars by comparing extinction and distance corrected $I$-band magnitudes with
the 4.5Myr isochrones of \cite{baraffe98}. For this we need the apparent distance modulus in $I$, which we calculate 
from the true distance modulus (Section~\ref{true_dist_mod}) and $A_I$. We do this using the ratio $A_I/E(B-V)_B$, 
calculated for red stars in the same way as described in Section~\ref{reddening}. We find this ratio is 1.94 (c.\,f. 1.89 
from \citealt{bessel98}), giving an apparent $I$-band distance modulus for Cep OB3b of 10.4. We also calculated 
masses for the periodic stars in NGC 2264 and NGC 2362. We adopted ages, true distance moduli and $E(B-V)_B$ from \cite{2008MNRAS.386..261M}. $A_I$ is calculated as outlined above to obtain the apparent distance modulus in $I$, and the masses are derived by comparison with the same \cite{baraffe98} isochrones
at the appropriate age. For the ONC a similar method was used, except that individual de-reddened $I$ band 
magnitudes of \cite{hillenbrand97} were used. In this way, we have ensured that all clusters have masses and ages 
which are determined in a self-consistent manner. The results are shown in Figure~\ref{fig:pdist_v_mass_all}, which  
shows the rotation periods of young stars in Cep OB3b, NGC 2362 NGC 2264 and the ONC, plotted as a function of 
mass.

%\begin{figure}
%\begin{center}
%\includegraphics[scale=0.35,angle=-90,trim=0 0 0 0,clip]{plots/cepPvM.ps} 
%\caption{Rotation period plotted against mass for the stars in Cepheus OB3b. See text for details.}
%\label{fig:pdist_v_mass}
%\end{center}
%\end{figure}
We can see from Figure~\ref{fig:pdist_v_mass_all}  that in many young clusters there is a strong dependence of 
rotation with mass, in the sense that the stars with masses below 0.4 M$_{\odot}$ show a dearth of slow rotators. This 
can readily be seen in the period distributions of NGC 2264 and NGC 2362, for example. To study this effect in Cep 
OB3b, we divide our data into high and low mass samples. The high mass sample contains stars with masses 
between 1.0 and 0.4 M$_{\odot}$, whilst the low mass sample contains stars with masses between 0.4 and 0.2 M
$_{\odot}$. The rotation distribution in Cep OB3b supports the general picture seen in other clusters; the low mass 
stars are rotating, on average, more rapidly than the higher mass stars. The median period of the low-mass stars is 3.7 days, against 4.4 days for the high mass stars. Unlike in other clusters, however, this result is 
barely statistically significant. A 1D K-S test gives a 6 per cent chance that the low and high mass stars were drawn 
from the same period distribution. Under the assumption that the period distributions in high and low-mass stars differ only in their median value, a Mann-Whitney test gives a 3 per-cent chance that the median period is the same for the high and low-mass stars in Cep OB3b. The difference is much more marked in other clusters; the same K-S test on the NGC 2362 sample gives a probability of less than 0.01 per cent that the high and low mass stars were drawn from the same distribution. Thus, whilst the stars in Cepheus show the same mass dependence of rotation seen elsewhere, the difference is less marked than is seen in other clusters.

\subsubsection{Comparison with NGC 2362}

It is apparent that the low-mass stars in Cep OB3b are rotating more slowly than those in other clusters. A comparison 
with NGC 2362 is most interesting, as it is at a similar age to Cep OB3b. The median period for low-mass stars in Cep 
OB3b is 3.7 days, compared to 2.3 days for NGC 2362. A 1D K-S test gives a 0.1 per cent chance that the periods of low mass stars in the two clusters were drawn from the same period distribution. Amongst the high mass stars the picture is different, with the high mass stars in Cep OB3b rotating more rapidly than their counterparts in NGC 2362. The median period for the  high-mass stars in Cep OB3b is 4.4 days, compared to 4.9 days for NGC 2362. This difference is only weakly significant though; a 1D K-S test gives a 6 per cent chance that the periods of high mass stars in the two clusters were drawn from the same period distribution. It is not clear why the two clusters should be so different.  Our simulations (see section~\ref{subsec:biasses}) suggest that we should have found almost all periods below 7 days with amplitudes greater than 0.02 mags in stars down to 0.2M$_{\odot}$. Similar calculations by \cite{irwin08} suggest that the survey in NGC 2362 was sensitive to periods of up to 10 days, with the same amplitude and for stars of the same mass. Thus, biasses could potentially have explained an  deficit of slow rotators in Cep OB3b, but not an excess. 
 
Another possibility is contamination of our periodic sample. We have taken steps to remove foreground and background contamination from our sample (see section~\ref{subsec:cleanup}), but it 
remains possible that there is contamination from a second population of PMS stars towards Cepheus OB3b. This 
second population need not share the distance and age of the main population, so masses assigned to its members 
would be in error. In this way slowly rotating, high mass PMS stars might be confused for lower mass objects. Optical 
spectroscopy of the region could test this possibility. If systematic error is not responsible for the observed difference 
between Cep OB3b and NGC 2362, it follows that different clusters at similar ages can show very different rotation 
distributions. This is not the first time such environmental differences have been found; the young stars in IC348 rotate 
much slower than those in the similarly aged NGC 2264 \citep{littlefair05}. 

\subsection{Rotation Period and Halpha emission}
\label{sec:p_v_disc}
\begin{figure}
\begin{center}
\includegraphics[scale=0.55,angle=0,trim=0 0 0 0,clip]{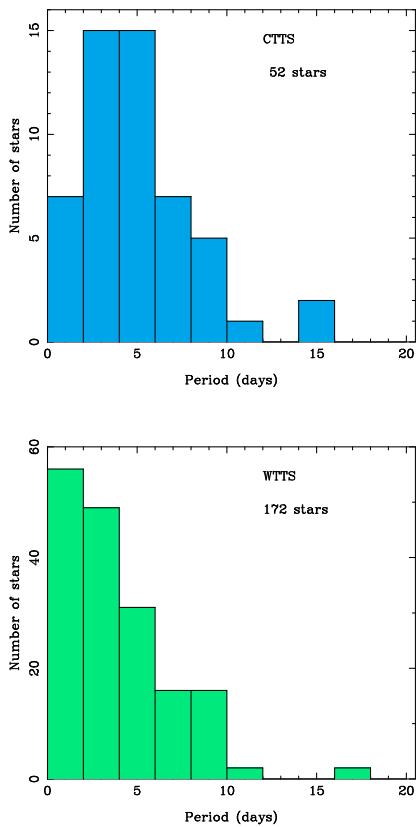} 
\caption{Period distributions for Cep OB3b, divided into samples
according to H$\alpha$ strength. The stars are grouped in accreting, classical T-Tauri stars (CTTS) and non-
accreting, Weak-lined T-Tauri stars (WTTS). See text for details of how stars were divided into these samples. }
\label{fig:pdist_v_disc}
\end{center}
\end{figure}

In order to gauge the effect of accretion upon rotation rate we examine here the link between rotation period and H$
\alpha$ emission. Other authors \citep[e.g.][]{rebull06,cieza07} have examined the link between mid-infrared excess 
and rotation. Whilst an infrared excess is good evidence for a disc it is not evidence for ongoing {\em accretion}, which 
is necessary for angular momentum regulation. H$\alpha$ emission, on the other hand, is a signature of ongoing 
accretion, and might also be expected to correlate with rotation rate. Indeed, \cite{lamm05} show that there is such a 
correlation in NGC 2264, in the sense that the stars showing strong H$\alpha$ emission rotate more slowly, as a 
group, than those lacking H$\alpha$ emission. A word of caution is needed: H$\alpha$ emission strength is only an 
indicator of {\em current} accretion rate. Since the star-disc interaction, or a stellar wind, can only remove angular 
momentum on a finite timescale it would be more meaningful to compare rotation rate against the {\em time-
averaged} accretion rate, taken over a comparable timescale. Additionally, it should always be born in mind that
photometric monitoring can in principle be biassed against the detection of periods in strongly accreting stars, although \cite{littlefair05} showed that dense temporal sampling, as used in this paper, is effective in reducing such a bias.

We divide our sample into accreting and non accreting objects in the $V-I$, $R-H\alpha$ diagram, following 
\cite{lamm04}.
Weak-line T-Tauri stars (WTTS) are assumed to be non-accreting, and are defined in this paper as objects with little or 
no H$\alpha$ emission, as defined by
\begin{equation}
(R-H\alpha) - (R-H\alpha)_{locus} + \Delta(R-H\alpha) < 0.0,
\end{equation}
where $(R-H\alpha)_{locus}$ is defined by equation~\ref{eq:locus}. Classical T-Tauri stars (CTTS) are objects with 
strong H$\alpha$ emission, and are assumed to be accreting. In this paper, we define CTTS as those objects which 
satisfy
\begin{equation}
(R-H\alpha) - (R-H\alpha)_{locus} - \Delta(R-H\alpha) > 0.1.
\end{equation}
\cite{lamm05} show that equation 4 corresponds roughly to $W_{\lambda}(H_{\alpha}) > 10 \AA$, which is frequently used to select CTTS, based on the equivalent width of H$\alpha$.

We find evidence that accretion is influencing rotation in Cepheus OB3b. The CTTS are rotating more 
slowly on average than the WTTS (see figure~\ref{fig:pdist_v_disc}. This difference is statistically  
significant; a 1D K-S test gives a 5 per cent chance they were drawn from the same parent distribution. Thus, the data 
in Cepheus lends some support to the generally accepted picture that accretion discs play a significant role in 
regulating the angular momentum of young stars. The result is only just statistically significant, however, and a 
statistically significant difference between CTTS and WTTS is no longer present if the data is broken down into high- 
and low-mass sub-samples. This is perhaps not surprising. Our method of dividing stars into CTTS and WTTS is 
rather crude, relying as it does on narrowband photometry, rather than spectroscopy; it is not without sensitivity 
biasses which can be a function of mass. Also, we classify relatively few objects as CTTS, which reduces the 
statistical power of our sample. To test the influence of present-day accretion on rotation periods, good quality H-$\alpha$ spectroscopy of our periodic objects is highly desirable.

%\section{Rotation Period and ``Age''}
%\label{sec:p_v_age}
%\begin{figure*}
%\begin{center}
%\includegraphics[scale=0.58,angle=0,trim=0 0 0 0,clip]{plots/Subgroups.ps} 
%\caption{}
%\label{fig:subgroups}
%\end{center}
%\end{figure*}

\section{Discussion}
\label{sec:discussion}

Our results show that Cepheus OB3b is a rich cluster for studying the evolution of stellar angular momentum. The 
periodic data in Cep OB3b broadly confirms the general picture of angular momentum evolution in young stars; 
accreting objects are rotating more slowly than non-accretors, confirming the influence of discs on the angular 
momentum of young stars. As seen in other young clusters, the low mass stars are rotating more rapidly than the high 
mass stars, but this mass dependence is much weaker than that typically observed. Compared to the similarly aged 
cluster NGC 2362, Cep OB3b has very slowly rotating low-mass stars, and possibly an excess of fast rotators amongst 
the high mass stars. There remains a possibility that this difference between the two clusters is a systematic error, 
perhaps related to contamination within our periodic database by a second PMS population. Optical spectroscopy of 
the region can test for this systematic effect, and can also improve our mass determinations, by determining individual 
extinction values for our periodic stars.

If this difference between the two clusters is not systematic in origin, it confirms the findings of \cite{littlefair05}, that 
clusters of similar ages can show very different rotational period distributions. Such environmental differences could possibly arise as a result of differing angular momentum budgets between clusters, or they might reflect differences in disc lifetimes between star forming regions. These results should act as a 
cautionary tale for models which aim to reproduce the angular momentum evolution of stars \cite[e.g.][]
{irwin07a,herbst02}. These models are based on the assumption that the rotation period distributions of different 
clusters can be assembled into an evolutionary sequence, but this assumption is broken if environmental differences 
between the clusters plays a larger role than evolutionary effects, due to magnetic wind braking, disc locking, or 
contraction towards the main sequence. It is thus important to fit such models to datasets which include rotation 
periods from more than one cluster at each age, to gauge the impacts of environmental differences between clusters.

\section{Conclusions}
\label{sec:concl}

We present a photometric study of $I$ band variability towards the young association Cepheus OB3b. The study is 
sensitive to periodic variability on timescales of less than a day, to more than 20 days. The result is a database of 704 
periodic variables in the field of Cep OB3b. A random inspection of 200 of these objects suggest that around 97 per 
cent of these periods are genuine. Colour cuts using $V$, $I$, $R$ and narrowband H$\alpha$ photometry
are used to reject contaminating objects, leaving 475 objects with measured rotation periods, which are very likely 
pre-main-sequence members of the Cep OB3b star forming region. 

We revise the distance and age to Cep OB3b, putting it on the consistent age and distance ladder of 
\cite{2008MNRAS.386..261M}. This yields a distance modulus of 8.8$\pm$0.2 mags, corresponding to a distance of 
580$\pm$60 pc, and an age of 4-5Myrs. For the purposes of this paper we therefore adopt an age of 4.5Myr.

The rotation period distribution confirms the general picture of rotational evolution in young stars, exhibiting both the 
correlation between accretion and rotation expected from disc locking, and the dependence of rotation upon mass 
that is seen in other star forming regions. However, this mass dependence is much weaker than seen in other 
regions. Comparison to the similarly aged NGC 2362 shows that the low-mass stars in Cep OB3b are rotating much 
more slowly.
This points to a possible link between star forming environment and rotation properties. Such a link would call into 
question 
models of stellar angular momentum evolution, which assume that associations can be assembled into an 
evolutionary sequence, thus ignoring environmental effects.

\section*{\sc Acknowledgements}
SPL is supported by an RCUK fellowship.  This research has made use of NASA's
Astrophysics Data System Bibliographic Services. Based on observations
made with the Isaac Newton Telescope operated on the island of La
Palma by the Isaac Newton Group in the Spanish Observatorio del Roque
de los Muchachos of the Instituto de Astrofisica de Canarias. We thank the many
INT observers who performed the observations which constitute the LB0 dataset for us.
The Faulkes Telescope Project is an educational and research arm of the Las Cumbres Observatory Global Telescope Network (LCOGTN).

\bibliographystyle{mn2e}
\bibliography{abbrev,refs,refs2,distance}

\end{document}